\documentclass[sigchi, screen]{acmart}

\copyrightyear{2023}
\acmYear{2023}
\setcopyright{rightsretained}
\acmConference[CHI EA '23]{Extended Abstracts of the 2023 CHI Conference on Human Factors in Computing Systems}{April 23--28, 2023}{Hamburg, Germany}
\acmBooktitle{Extended Abstracts of the 2023 CHI Conference on Human Factors in Computing Systems (CHI EA '23), April 23--28, 2023, Hamburg, Germany}
\acmDOI{10.1145/3544549.3583178}
\acmISBN{978-1-4503-9422-2/23/04}

\begin{document}

\title{Human-Centered Responsible Artificial Intelligence: Current~\&~Future~Trends}

\author{Mohammad Tahaei}
\orcid{0000-0001-9666-2663}
\affiliation{
  \institution{Nokia Bell Labs}
  \city{Cambridge}
  \country{UK}
}
\email{mohammad.tahaei@nokia-bell-labs.com}

\author{Marios Constantinides}
\orcid{0000-0003-1454-0641}
\affiliation{
  \institution{Nokia Bell Labs}
  \city{Cambridge}
  \country{UK}
}
\email{marios.constantinides@nokia-bell-labs.com}

\author{Daniele Quercia}
\orcid{0000-0001-9461-5804}
\affiliation{
  \institution{Nokia Bell Labs}
  \city{Cambridge}
  \country{UK}
}
\email{daniele.quercia@nokia-bell-labs.com}

\author{Sean Kennedy}
\orcid{0000-0003-0000-1786}
\affiliation{
  \institution{Nokia Bell Labs}
  \city{Ottawa}
  \country{Canada}
}
\email{sean.kennedy@nokia-bell-labs.com}

\author{Michael Muller}
\orcid{0000-0001-7860-163X}
\affiliation{
    \institution{IBM Research AI}
    \city{Cambridge}
    \state{MA}
    \country{USA}
}
\email{michael_muller@us.ibm.com}

\author{Simone Stumpf}
\orcid{0000-0001-6482-1973}
\affiliation{
    \institution{University of Glasgow}
    \city{Glasgow}
    \country{UK}
}
\email{simone.stumpf@glasgow.ac.uk}

\author{Q. Vera Liao}
\orcid{0000-0003-4543-7196}
\affiliation{
    \institution{Microsoft Research}
    \city{Montreal}
    \country{Canada}
}
\email{veraliao@microsoft.com}

\author{Ricardo Baeza-Yates}
\orcid{0000-0003-3208-9778}
\affiliation{
    \institution{EAI, Northeastern University}
    \state{CA}
    \country{USA}
}
\email{rbaeza@acm.org}

\author{Lora Aroyo}
\orcid{0000-0001-9402-1133}
\affiliation{
    \institution{Google}
    \state{NY}
    \country{USA}
}
\email{l.m.aroyo@gmail.com}

\author{Jess Holbrook}
\orcid{0000-0001-9398-6056}
\affiliation{
    \institution{Meta}
    \city{}
    \country{USA}
}
\email{jess.holbrook@gmail.com}

\author{Ewa Luger}
\orcid{0000-0001-7882-9415}
\affiliation{
    \institution{University of Edinburgh}
    \city{Edinburgh}
    \country{UK}
}
\email{ewa.luger@ed.ac.uk}

\author{Michael Madaio}
\orcid{0000-0003-2133-0810}
\affiliation{
    \institution{Google}
    \city{}
    \country{USA}
}
\email{madaiom@google.com}

\author{Ilana Golbin Blumenfeld}
\orcid{0000-0001-8366-8468}
\affiliation{
    \institution{PwC}
    \city{Los Angeles}
    \state{CA}
    \country{USA}
}
\email{ilana.a.golbin@pwc.com}

\author{Maria De-Arteaga}
\orcid{0000-0003-2297-3308}
\affiliation{
    \institution{University of Texas at Austin}
    \state{TX}
    \country{USA}
}
\email{dearteaga@mccombs.utexas.edu}

\author{Jessica Vitak}
\orcid{0000-0001-9362-9032}
\affiliation{
    \institution{University of Maryland, College Park}
    \state{MD}
    \country{USA}
}
\email{jvitak@umd.edu}

\author{Alexandra Olteanu}
\orcid{0000-0001-5710-4511}
\affiliation{
    \institution{Microsoft Research}
    \state{Montreal}
    \country{Canada}
}
\email{alexandra.olteanu@microsoft.com}

\renewcommand{\shortauthors}{Tahaei et al.}

\begin{abstract}

In recent years, the CHI community has seen significant growth in research on \textit{Human-Centered Responsible Artificial Intelligence}. While different research communities may use different terminology to discuss similar topics, all of this work is ultimately aimed at developing AI that benefits humanity while being grounded in human rights and ethics, and reducing the potential harms of AI. In this special interest group, we aim to bring together researchers from academia and industry interested in these topics to map current and future research trends to advance this important area of research by fostering collaboration and sharing ideas.

\end{abstract}



\maketitle

\section{Motivation \& Background}
\textit{Human-Centered Responsible Artificial Intelligence (HCR-AI)}\footnote{Different communities have adopted different terminologies to address related topics. We intentionally left the proposal and terminology open without emphasizing specific topics to attract participants from various backgrounds and interests. One reason to propose this SIG is to discuss various aspects of HCR-AI with researchers who can provide diverse perspectives.} aims to bring people and their values into the design and development of AI systems, which can contribute to building systems that benefit people and societies, as well as preventing and mitigating potential harms. Despite a long history of the importance of the human factor in AI systems~\cite{friedman1966bias, suchman1987plans}, there has been a growing awareness of its importance within the CHI community in the past few years~\cite{tahaei2023toward}. Searching the ACM Digital Library within CHI proceedings shows the following results (Figure~\ref{fig:trends}):\footnote{Using search within anywhere on the ACM Digital Library. Results are not mutually exclusive and include all types of materials (e.g., research papers, extended abstracts, panels, and invited talks). Filtering for only research papers results in 32 unique papers since 2020. We acknowledge this is not an exhaustive search and is only to show the growing body of research in CHI.} ``human-centered AI'' results in 41 records since 2019 and ``responsible AI'' results in 32 records since 2020. Below, we highlight a few examples of these studies, which are relevant to the topic of the Special Interest Group (SIG), noting that this is not an exhaustive list and is only to show the breadth and depth of the existing work: 

\begin{figure*}
  \centering
  \includegraphics[width=.8\textwidth]{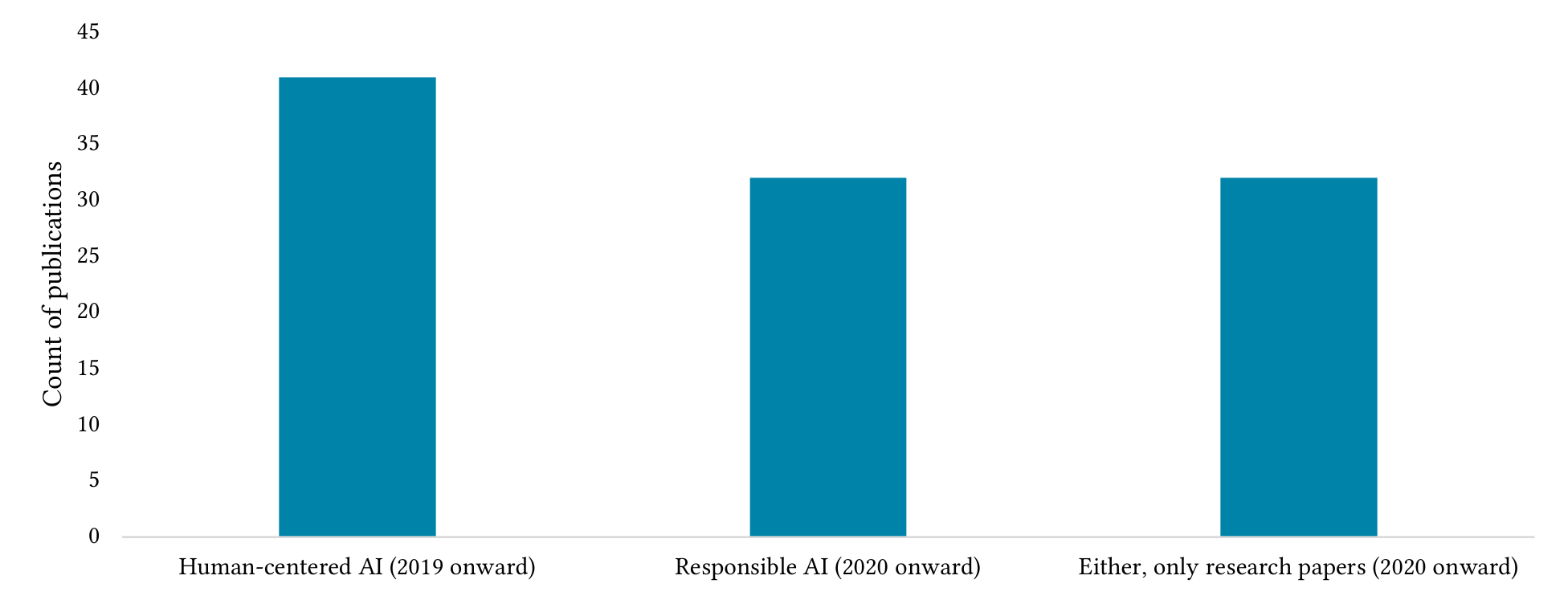}
  \caption{Counts of publications containing ``human-centered AI'' and ``responsible AI'' at CHI. The first three bars for human-centered AI and responsible AI are not mutually exclusive. They include all types of materials (e.g., research papers, extended abstracts, and invited talks). Filtering for only research papers results in 32 unique papers since 2020 (the last bar).}
  \label{fig:trends}
\end{figure*}

Ethics in AI involve socio-cultural and technical factors, spanning a range of responsible AI values (including but not limited to transparency, fairness, explainability, accountability, autonomy, sustainability, and trust)~\cite{jobin2019global}. However, different stakeholders, including the general population and AI practitioners, may perceive and prioritize these values differently. For example, a representative sample of the U.S. population was more likely to value safety, privacy, and performance. In contrast, practitioners were more likely to prioritize fairness, dignity, and inclusiveness~\cite{jakesch2022different}. Or, certain historically exploited groups may weigh privacy or non-participation more highly than groups with lower risk~\cite{garcia2020no, muller2022forgetting}.

Aligned with responsible AI are calls to make AI more human-centric. In particular, there is an emphasis on the challenges of AI integration into socio-technical processes to preserve human autonomy and control, as well as the impacts of AI systems deployment and applications on society, organizations, and individuals~\cite{boyarskaya2020overcoming}. On this strand of research, understanding \textit{socio-technical} and \textit{environmental} factors can help surface why and how an AI system may become human-centered~\cite{shneiderman2020bridging, ehsan2021expanding, liao2021human}. For example, even for an AI for which there might be broader consensus on its utility, such as the detection of diabetes using retina scans, there may well be barriers to becoming useful for its intended users, including due to not fitting well with the users' workflows (e.g., nurses) or the system requiring high-quality images that are not easy to produce, especially in locations with low resources where such technology can provide significant support to patients if done right~\cite{beede2020human}.

Similarly, researchers have looked at individuals' expectations and understandings of AI. For example, when making an \textit{ethical decision} (e.g., a hypothetical scenario for bringing down a terrorist drone to save lives), people may put more \textit{capability trust} in an AI decision maker (i.e., capacity trustworthiness, being more capable), whereas they may put more \textit{moral trust} in a human expert (i.e., being able to be morally trustworthy and make decisions that are aligned with moral values); in either case, decision made by a human or an AI, prior work has found that people often see the human as partly responsible, be it the decision maker or the AI developer~\cite{tolmeijer2022capable}---even though the outcomes of the developer may intentionally or unintentionally limit the span of action of the decision-maker~\cite{muller2022frameworks}. Regarding moral dilemmas between AI and human decisions, people may not equally judge humans and machines~\cite{hidalgo2021humans}. These variations in perceptions may be rooted in (a) people judging humans by their intentions and machines by their outcomes, and (b) people assigning extreme intentions to humans and narrow intentions to machines, while they may excuse human actions more than machine actions in accidental scenarios~\cite{hidalgo2021humans}. Furthermore, people's perceived fairness and trust in an AI may change with the terminology used to describe it (e.g., an algorithm, computer program, or artificial intelligence), which could eventually impact the system's success and outcomes, especially when comparative research is done~\cite{langer2022look}. 

Another human aspect of AI systems is the people who work on these systems, such as annotators, engineers, and researchers. Data annotators are part of the workforce that produces the datasets used to train AI models. However, the workforce (sometimes referred to as \textit{AI labor}~\cite{crawford2022atlas} or \textit{ghostworkers}~\cite{gray2019ghost}) behind the annotation task may have career aspirations that the current annotation companies do not support, or they may be poorly paid because of the push that comes from the recent development in AI that requires massive annotated datasets at low costs~\cite{wang2022whose, gould2022consumption}. Other researchers echo similar observations about AI labor by saying that ``without the work and labor that were poured into the data annotation process, ML [Machine Learning] efforts are no more than sandcastles,''~\cite{wang2022whose} or ``everyone wants to do the model work, not the data work,''~\cite{sambasivan2021everyone} a behavior that contributes to the creation of \textit{data cascades}---which refer to compounding events causing adverse, downstream effects from data issues, resulting in technical debt.\footnote{In 1992, Ward Cunningham put forward the metaphor of technical debt to describe the build-up of cruft (deficiencies in internal quality) in software systems as debt accrual, similar to financial~\cite{cunningham1992wycash} or ethical debt~\cite{wired_ethical_debt} (i.e., ``AI ethical debt is incurred when an agency opts to design, develop, deploy and use an AI solution without proactively identifying potential ethical concerns''~\cite{petrozzino2021pays}).}

New tools and frameworks are now being proposed to help developers build more responsible AI systems (e.g., IBM's 360 suites on fairness and explainability~\cite{ibm2019fairness,ibm2022ai} and Fairlearn~\cite{bird2020fairlearn}), in addition to user-led approaches to algorithmic auditing to uncover potential harms of algorithmic systems~\cite{devos2022toward}. Despite the growing interest in HCI research and user experience design for AI, developing responsible AI remains challenging; a mission involving cognitive, socio-technical, cultural, and design perspectives~\cite{liao2021human, gunning2019xai, lee2021included}.

These are just a few examples from many studies that cover topics that have emerged within the past few years and are relevant to the SIG's scope. Besides CHI, the ACM Conference on Fairness, Accountability, and Transparency (\textit{ACM FAccT}), established in 2018~\cite{facct2022}, aims to bring ``together researchers and practitioners interested in fairness, accountability, and transparency in socio-technical systems'' highlighting the importance of the research in HCR-AI. We aim to bring this community together in a 75-minute discussion and brainstorming session at CHI 2023.

\section{Proposal \& SIG's Goal}
The SIG follows similar strands from past workshops at CHI 2020, 2021, and 2022~\cite{lee2020human, ehsan2021operationalizing, ehsan2022human}. The topics discussed are evolving and growing (Figure~\ref {fig:trends}); hence, a SIG at CHI 2023 would be timely. We believe a SIG dedicated to the HCR-AI at CHI 2023 will benefit the CHI community and help build and establish a broader network of researchers and provide a mapping and understanding of current and future trends in this area. Researchers in this area come from industry and academia from diverse disciplinary backgrounds (e.g., theoretical computer science, social computing, machine learning, human-computer interaction, and social science). Therefore, having them all in one hybrid physical-virtual room for 75 minutes would benefit the community and the attendees to brainstorm and generate a map of current and future trends in this area (activity diagramming). We propose to use online tools such as Miro and Slack to (a) create a record of the group's co-constructed knowledge; (b) serve as a persistent communication to others in the CHI community; and (c) enfranchise remote participants.

\section{Expected Outcomes \& Next Steps}
We will share the Miro board with attendees and make it public to support future research in HCR-AI. We will also create a Slack channel for future communications. The SIG's primary goal is to create a sense of community among researchers in this area, from academia and industry, to establish collaborations. The SIG is an excellent opportunity to bring people with a shared interest in HCR-AI who also attend CHI to build this community.

After the SIG, we will organize virtual biannual meetings with the attendees to share their latest ideas and recent work, build a website to share outcomes created during the SIG, encourage attendees to apply for joint grants, and explore the possibility of creating a symposium similar to CHIWORK.

\newpage

\balance
\bibliographystyle{ACM-Reference-Format}
\bibliography{biblio}

\appendix

\clearpage
\thispagestyle{empty}

\section*{Supplementary Material for Human-Centered Responsible AI: Current \& Future Trends}

\subsection*{Target Community \& Attendees}
We expect attendees to have experience in either Human-Computer Interaction (HCI) or Artificial Intelligence (AI), broadly defined, with a shared interest in AI's ethics and the human factor. Attendees will bring experience from the technical side of AI as well as the social and human side of HCI. The interaction and discussion between the two groups is the goal of this SIG. We expect to see attendees from academia and industry, as the research in responsible AI is taking place in both communities.

We will use social media (e.g., Mastodon, Twitter, LinkedIn, and Facebook CHI Meta) to publicize the SIG. We will also use our network to attract more participants, as the authors all have experience in similar domains. Organizers from various institutes and backgrounds share an interest in the SIG's topic, enabling us to advertise the SIG in different communities. We will promote the SIG for attendees from a wide range of career levels, from students interested in the topic to senior individuals with years of experience.

\subsection*{Presentation \& Schedule}
We will use Miro, an online collaboration tool, to facilitate the discussion in the SIG. Below, we propose our schedule:

\begin{itemize}
    \item Introduction (10 minutes): organizers will briefly introduce themselves and the SIG. Depending on the number of attendees, we may do a quick round table with short introductions.
    \item Group activity (20 minutes): attendees will form groups of four and work on a Miro board. During this activity, they will discuss current trends and what is needed in the future.
    \item Presentation (15 minutes): each group will present their findings for three minutes to the rest of the attendees.
    \item Merge and discuss (20 minutes): all attendees will merge all the findings in one board, removing duplicates, adding labels to emerging themes, and finding relations between the themes.
    \item Final discussion (10 minutes): organizers will facilitate discussion around the final board. The goal is to leave the SIG with a clean mapping of current and future trends related to the SIG's topic.
    \item Lunch or dinner: we will encourage attendees to join for a group lunch or dinner after the SIG (paid by the attendee).
\end{itemize}

\subsection*{Contacts}
Mohammad Tahaei and Marios Constantinides from Nokia Bell Labs, Cambridge, UK, will be the first point of contact.

\end{document}